\documentclass[aps,pre,preprint,showpacs,showkeys,superscriptaddress]{revtex4-1} 
\usepackage[utf8]{inputenc}
\usepackage{cmap} 
\usepackage[T1]{fontenc}
\usepackage{microtype}
\usepackage{amsmath,amssymb,mathtools}
\usepackage{txfonts}

\usepackage{mathrsfs}
\usepackage{graphicx,grffile,textcomp}
\graphicspath{{figure/}}

\usepackage{textcomp}
\usepackage{booktabs,tabularx,dcolumn}
\newcolumntype{d}[1]{D{.}{.}{#1}}

\usepackage{url,hyperref}
\usepackage[usenames,dvipsnames]{xcolor}
\hypersetup{colorlinks=true, linkcolor=BrickRed, urlcolor=blue!50!black, citecolor=blue!50!black}

\usepackage[capitalize]{cleveref}
\crefrangelabelformat{equation}{(#3#1#4)$-$(#5#2#6)}

\usepackage{placeins}

\begin{document}
\title{Flux and storage of energy in non-equilibrium, stationary states}

\author{Robert Holyst$^*$}
\affiliation{Institute of Physical Chemistry, Polish Academy of Sciences, 
Kasprzaka 44/52, PL-01-224 Warsaw, Poland} 
  
\author{Anna Macio\l ek}
\affiliation{Institute of Physical Chemistry, Polish Academy of Sciences, 
Kasprzaka 44/52, PL-01-224 Warsaw, Poland} 

\author{Yirui Zhang }
\affiliation{Institute of Physical Chemistry, Polish Academy of Sciences, 
Kasprzaka 44/52, PL-01-224 Warsaw, Poland} 

\author{Marek Litniewski}
\affiliation{Institute of Physical Chemistry, Polish Academy of Sciences, 
Kasprzaka 44/52, PL-01-224 Warsaw, Poland} 

\author{Piotr Knycha\l a}
\affiliation{The President Stanisław Wojciechowski State University of Applied Sciences, Nowy Świat 4, PL-62-800 Kalisz, Poland}

\author{Maciej Kasprzak}
\affiliation{Adam Mickiewicz University, Faculty of Physics and NanoBioMedical Centre, Umultowska 85, PL-61-614, Poznan, Poland}

\author{Micha{\l}  Banaszak}
\affiliation{Adam Mickiewicz University, Faculty of Physics and NanoBioMedical Centre, Umultowska 85, PL-61-614, Poznan, Poland}
\date{\today}
\begin{abstract}
Systems kept out of equilibrium in stationary states  by an external source of energy store  an energy $\Delta U=U-U_0$. $U_0$ is the internal energy at equilibrium state, obtained after the shutdown of energy input. We determine $\Delta U$  for two model systems: ideal gas and Lennard-Jones fluid. $\Delta U$ depends not only on the total energy flux, $J_U$,  but also on the mode of energy transfer into the system.  We use three different modes of energy transfer where: the energy flux per unit volume is (i) constant; (ii)  proportional to the local temperature (iii) proportional to the local density. We show that  $\Delta U /J_U=\tau$ is minimized in the stationary states formed in these systems, irrespective of the mode of energy transfer. $\tau$ is the characteristic time scale of energy outflow from the system immediately after the shutdown of energy flux. We prove that $\tau$ is minimized in stable states of the Rayleigh-Benard cell. 
\end{abstract}
\maketitle
Systems out of equilibrium are notoriously difficult to describe in  a single coherent methodology based on variational principles. Principles such as Prigogine minimum entropy production~\cite{jaynes1980}, Attard second entropy variation~\cite{attard2008} or, Ziegler maximum entropy production~\cite{martyushev2013} etc. suggested over the last 100 years, have not reached the same status as the maximum entropy principle known from equilibrium thermodynamics~\cite{vita2010,bartlett2016,velasco2011}. A new paradigm, such as the driven lattice gas system, is believed to become an "Ising model" for non-equilibrium statistical physics~\cite{katz1984,zia2010,stinchcombe2001,pleimling2010,dickman2018,zia2002}.  Steady State Thermodynamics (SST) is yet another description framework for  non-equilibrium stationary  states, which is still being developed~\cite{dickman2014,pradhan2011,sasa2006,oono1998}.  Here we present a different approach to stationary states, based on two quantities: the energy stored 
in non-equilibriums states,  $\Delta U$, and the total energy flux, $J_U$ in these states. 

 The second law of thermodynamics states that the entropy of a system has its maximum value at the equilibrium state. Entropy, $S$,  is a function of state, thus  for an isolated system of $N$ molecules of total internal energy $U$ enclosed in a volume, $V$ , the entropy has a fixed value $S=S(U,V,N)$. 
Internal constraints make it possible to divide the system into $n$ isolated subsystems of entropies $S_i=S(U_i,V_i,N_i)$ for $i=1\cdots n$ , where $\sum_i^n N_i=N$, $\sum_i^n V_i=V$, $\sum_i^n U_i=U$. The maximum entropy principle states~\cite{callen1985,holyst2012} that $S(U,V,N)\ge \sum_i^n S_i$.   We develop a similar methodology for non-equilibrium stationary states, by introducing internal constraints in these states. Next we make  a conjecture on  variational principles for these states.  We do not claim the generality of this conjecture, but simply prove it for three studied cases. In order to illustrate our methodology we study two systems: ideal gas and Lennard-Jones fluid. Next we test our methodology on two competing states in the Rayleigh-Benard (RB) cell~\cite{cross1993}. Summarizing our results obtained for these systems: we find that  $\Delta U$ is minimized in non-equilibrium stationary states for a fixed $J_U$. Minimization is with respect to all constrained states of the system, similarly as in equilibrium thermodynamics. Due to the lack of a general theoretical framework for describing  the non-equilibrium systems under consideration, we are not able to formulate a general argument for the validity of the conjectured principle. Nevertheless, it is plausible that, under  well defined conditions,  the energy storage  obeys some sort of  variational  principle. In the search for this principle  we employed  dimensional analysis and considered the relevant physical quantities. This analysis suggests the ratio $\Delta U/J_U$,  which has the dimension of time and a nice physical interpretation, as a  candidate for the quantity to be extremized.

{\it Ideal gas between two planar walls}:  An  ideal gas is confined between two planar walls of  surface area $\mathcal{A}$  located at $z=\pm L$.
The temperature of the  walls is constant $T_{-L}=T_{L}=T_0$. 
 Energy is supplied to the volume of the fluid (e.g. by microwaves). In stationary states the total flux of energy into the system, $J_U$, matches the total flux at the walls.
In the hydrodynamic limit, the time evolution of the system is given by the conservation laws  for mass, momentum, and energy 
supplemented  by the relations between thermodynamic forces and fluxes and thermodynamic equations of state. For an ideal gas
$U=(3/2)Nk_BT$, $p=\rho k_BT$, where $U$ is the internal energy, $p$ is the pressure, $\rho=N/V$ is the number density of a fluid, $V$ is the volume, $N$ is the number of gas particles and $k_B$ is the Boltzmann constant. We observed in our previous simulations/calculations~\cite{babin2005,holyst2008} for gas-liquid evaporating systems that mechanical equilibrium is established very fast (in comparison to heat flow). Therefore we also expect a constant pressure across the system in this case.
The stationary state 
 satisfies $\mathbf{v}=0$ (gas velocity), $p=const$~\cite{babin2005,holyst2008}. These two equations come from the conservation of mass and momentum. 
 { (The same results, i.e., $\mathbf{v}=0$ and $p=const$ in the stationary states are also obtained in MD simulations  of the Lennard-Jones system described later.)} 
 The conservation of energy is given by
$\nabla \cdot \mathbf{j}_{\varepsilon}(\mathbf{r})=\sigma_{\varepsilon} $, where $\mathbf{j}_{\varepsilon}(\mathbf{r})$ is local heat flux and $\sigma_{\varepsilon} $
is the external  heat source. The integral of $\sigma_{\varepsilon} $ over the volume is $\int_V\sigma_{\varepsilon}= J_U$. The local flux of energy per unit area is proportional to the temperature gradient $\mathbf{j}_{\varepsilon} = -\kappa \nabla T$, where $\kappa$ is the thermal conductivity and thus:
\begin{equation}
 \label{eq:1}
  -\kappa \nabla^2 T(\mathbf{r}) = \sigma_{\varepsilon}.
\end{equation}
We solve Eq(\ref{eq:1}) for three  forms of the source term describing different manners of energy supply: (i)  $\sigma_{\varepsilon}=\lambda_1$, (ii)  $\sigma_{\varepsilon}=\lambda_2T(z)$, and (iii) $\sigma_{\varepsilon}=\lambda_3\rho(z)$.
The equations are transformed into dimensionless form by rescaling the variables  $\tilde z =z/L$, $\tilde T=T/T_0 $ and $\tilde\rho=\rho/\rho_0 $  together with (i) $\tilde\lambda_1 = \lambda_1 L^2/(\kappa T_0) $, (ii) $\tilde\lambda_2 = \lambda_2  L^2/\kappa$, and 
(iii) $\tilde\lambda_3 = \lambda_3L^2\rho_0/(\kappa T_0)$, { with $\lambda_i > 0, i=1,2,3$}.
Here, $\rho_0$ is the number density of the ideal gas at temperature $T_0$ and pressure $p_0$. 
The temperature is a function of coordinate $z$, only. In all these cases we introduce energy directly into the volume. This mode of energy transfer is well known by e.g. electromagnetic waves (microwaves , visible light etc.). If we couple a microwave device with a thermovision camera we can in principle add energy to hotter places (where density is low) or to less hot spots, where density is high. Thus the three modes of energy transfer are physically feasible. 

For case (i) the equation for stationary temperature profile is
$d^2 \tilde T(\tilde z)/d\tilde z^2 = - \tilde\lambda_1$ with symmetric boundary conditions $\tilde T(-1)=\tilde T(1)=1$. This equation has the solution
$\tilde T(\tilde z)=-\frac{\tilde\lambda_1}{2}(\tilde z^2-1)+1$. The energy of this stationary state is calculated using  the assumption of local equilibrium. 
 The energy density field $U/(V\rho_0T_0)=\varepsilon(\tilde T(\tilde z),\tilde\rho(\tilde z))=(3/2)k_B\tilde\rho(\tilde z)\tilde T(\tilde z)$ for an ideal gas.
Upon rescaling by the equilibrium value $U_0/(V\rho_0 T_0)=\varepsilon_0=(3/2)k_B$, the dimensionless local energy density $\tilde\varepsilon=\varepsilon/\varepsilon_0$ obeys
$\tilde\varepsilon(\tilde T(\tilde z),\tilde \rho(\tilde z))/\tilde T(\tilde z)=\tilde \rho(\tilde z)$.  Our system does not exchange molecules with the environment therefore the number of particles is constant. This condition is given by the equation 
$\int_{-1}^1\tilde \rho(\tilde z)d\tilde z=2$ and thus implies
$\int_{-1}^{1}d\tilde z\tilde\varepsilon(\tilde T(\tilde z),\tilde \rho(\tilde z))/\tilde T(\tilde z)=2$.
The reduced  pressure obeys $\tilde p=\tilde T(\tilde z)\tilde \rho(\tilde z)=\tilde\varepsilon(\tilde T(\tilde z),\tilde \rho(\tilde z))$, thus since
$\tilde p$  is constant, so is  $\tilde \varepsilon$.  
We obtain $\tilde\varepsilon$  as
\begin{equation}
 \label{eq:2}
  \tilde\varepsilon = \frac{2}{\int_{-1}^{1}d\tilde z/\tilde T(\tilde z)}.
\end{equation}
Integrating the temperature profile we obtain 
\begin{equation}
 \label{eq:3}
  \tilde\varepsilon(\tilde \lambda_1) = \frac{\sqrt{\tilde\lambda_1(\tilde\lambda_1+2)}}{2\mathrm{tanh}^{-1}\Bigg(\sqrt{\frac{\tilde \lambda_1}{\tilde \lambda_1+2}}\Bigg)}.
\end{equation}
For  cases (ii) and (iii)   of
 the external flux see Supplemental Material  (SM). 

We introduce a rigid, impenetrable, adiabatic wall in the system  at $z=z_1$. The wall divides the system into two subsystems (1) and (2). 
The internal energy at equilibrium is $U_0=2L\mathcal{A}\varepsilon_0$, the same as in the unconstrained system.  
We calculate the storage of energy over its equilibrium value, $\Delta U_1+\Delta U_2= U_1+U_2-U_0$ in the same way as presented above. The division into two subsystems satisfies the following conditions in our methodology: 1) The subsystems are in mutual equilibrium after the shutdown of energy  input, so that no additional fluxes appear after removal of the constraints at equilibrium. 2) The subsystems reach a stationary state characterized by $U_1$, $U_2$ and fluxes $J_{U_1}$, $J_{U_2}$.  3) The mode of energy transfer to each subsystem is the same.
Since $J_U=\mathcal{A}\int_{-L}^{L}\sigma_{\varepsilon}dz$ we
obtain for case (i):
\begin{equation}
 \label{eq:4}
 \frac{\Delta U}{J_U}=\frac{\varepsilon_0}{\lambda_1}\Big(\tilde \varepsilon-1\Big)
 \end{equation}
 and 
 \begin{equation}
  \label{eq:5}
 \frac{\Delta U_1+\Delta U_2-U_0}{J_{U_1}+J_{U_2}}= \frac{\varepsilon_0}{\lambda_1}\Big(f_1(\tilde z_1)+f_2(\tilde z_1)-1\Big)
 \end{equation}
with 
\begin{equation}
 \label{eq:6}
 f_1(\tilde z_1) = \frac{(1-\tilde z_1)\sqrt{\Lambda^{(1)}_{-}(\Lambda^{(1)}_-+2)}}{4\mathrm{tanh}^{-1}\Big(\sqrt{\frac{\Lambda^{(1)}_-}{\Lambda^{(1)}_-+2}}\Big)}  \qquad \mathrm{and} \qquad   f_2(\tilde z_1) = \frac{(1+\tilde z_1)\sqrt{\Lambda^{(1)}_+(\Lambda^{(1)}_+ + 2)}}{4\mathrm{tanh}^{-1}\Big(\sqrt{\frac{\Lambda^{(1)}_+}{\Lambda^{(1)}_+ + 2}}\Big)}
\end{equation}
where $\tilde z_1=L_1/L$, $\Lambda^{(1)}_-=\tilde\lambda_1 (1 - \tilde z_1)^2$ and $\Lambda^{(1)}_+=\tilde\lambda_1 (1 + \tilde z_1)^2$.
By inspection, in the range   $0\le \tilde z_1 <1$ the functions $ f_i(\tilde z_1), i=1,2$ are positive and lie above their tangent lines at $\tilde z_1=0$, i.e.,
 $ f_i(\tilde z_1) > f'_i(\tilde z_1)\big|_{\tilde z_1=0} \tilde z_1 + \frac{\varepsilon}{2}, \; i=1,2$ for  fixed $\tilde\lambda_1 >0$.
 Since $f_1(\tilde z_1)=f_2(-\tilde z_1)$ one has $f'_1(\tilde z_1)\big|_{\tilde z_1=0} = - f'_2(\tilde z_1)\big|_{\tilde z_1=0}$ and hence 
$f_1(\tilde z_1)+f_2(\tilde z_1) > 2 \frac{\varepsilon}{2}$ which proves that 
\begin{equation}
 \label{eq:7}
 \frac{\Delta U_1+\Delta U_2}{J_{U_1}+J_{U_2}}\ge\frac{\Delta U}{J_U}.
\end{equation}
The equality holds only for equal partition of the system into two subsystems ($z_1=0$). This equation states that the energy stored in two subsystems is larger than in the unconstrained system for a fixed flux $J_U=J_{U_1}+J_{U_2}$.
This observation holds irrespective of the sign of $\Delta U$ i.e. irrespective of the equilibrium reference state {\cite{footnote}}.
Cases (ii) and (iii) also satisfy Eq.~(\ref{eq:7}) (discussed in SM). Additionally in SM we present calculations of case (i) with $\kappa={\rm const} \sqrt{T}$ (expected for the dilute gas){, which} further confirm Eq.~(\ref{eq:7}) for this system. It is well understood that a  true ideal gas would have non-interacting particles with a zero collision cross section. {We} use
the equation of state of the ideal gas as an
approximation for the interacting gas characterized by finite, temperature dependent, $\kappa$.

{\it Lennard-Jones liquid in a rectangular box:}
 In order to perform analytical calculations for the ideal gas model we had to assume local equilibrium. This assumption is inherent  in irreversible thermodynamics equations. In order to test our analytical { results} for the ideal gas model we performed Molecular Dynamics (MD) simulations of Lennard-Jones system. MD simulations provided qualitatively the same results as analytical calculations presented in the previous section. In the MD simulations Newton equations of motion are solved, thus { no assumptions concerning local equilibirum or constancy of heat conductivity are made}. {Nevertheless,  during MD simulations our system stays quite close to local equilibrium.} In general we can expect a violation of the local equilibrium assumption only when the flux of the energy flowing across the system is faster than the process of local distribution of energy between all degrees of freedom. Such conditions are expected in e.g. shock waves.

\begin{figure}[ht]
\includegraphics[width=1\textwidth]{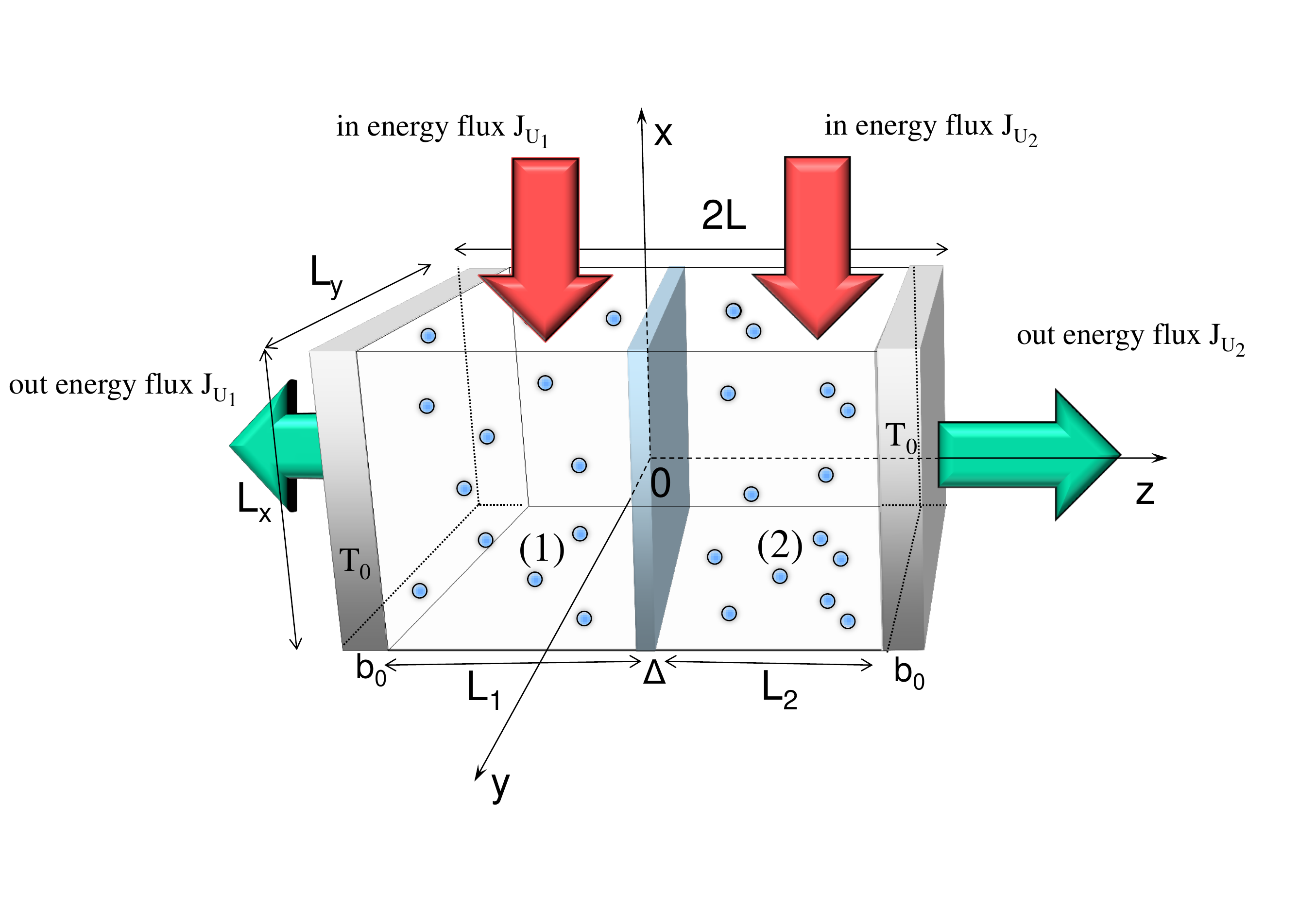} 	
\caption{  Lennard-Jones  simulation box with the shaded region in the middle dividing the system into (1) and (2) subsystems of size $L_1$ and $L_2$ respectively. The finite size of this wall precludes interactions between the molecules in different subsystems. Heating is indicated by the red arrows. Cooling is performed at the boundaries at  a distance $b_0$ from two walls and shown schematically by the green arrows. The fluxes $J_{U_1}$ and $J_{U_2}$ in and out are equal in the stationary state. The total size of the system $L$ is fixed. }
\label{fig1}
\end{figure}

\begin{figure}[ht]
\includegraphics[width=0.7\textwidth]{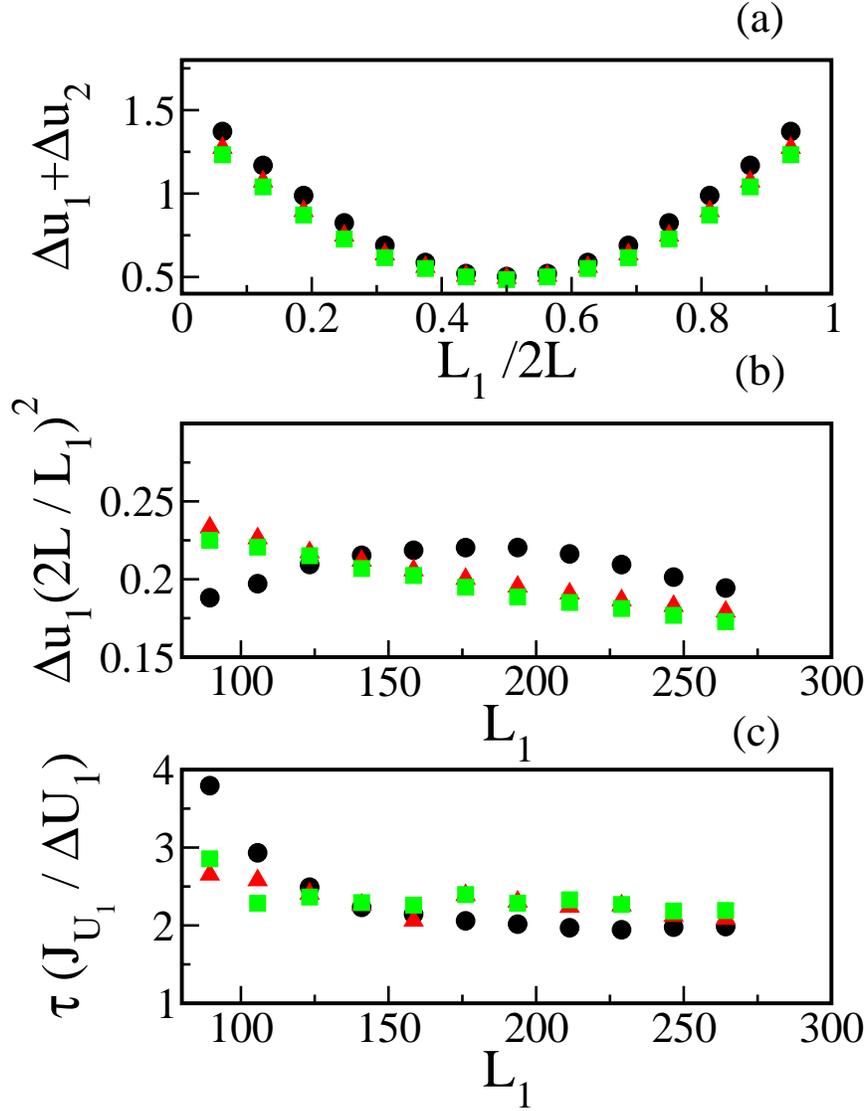} 	
\caption{ (a) Energy stored $\Delta u_1+\Delta u_2$  per LJ atom in subsystem (1) and (2) as a function of the size of one subsystem $L_1$ for a fixed flux $J_U=J_{U_1}+J_{U_2}$. The flux of energy per unit volume is constant (red triagles); proportional to temperature (black circles); proportional to density(green squares). The minimal value of $\Delta U_1+\Delta U_2=\Delta U$  is equal to the value for the unconstrained system for $L_1=L_2$ (see Eq(\ref{eq:7})). (b) The energy (per particle) in subsystem (1) as a function of $L_1$. This energy scales as $U_1/V\sim L_1^2$. It is not an extensive quantity, since the energy at equilibrium $U_0/V\sim {const}$.
(c) The characteristic time $\tau$ of energy out-flow from the sub-system (1) after the shutdown of energy flux into the sub-system. Energy decreases as $\exp(-t/\tau)$ (see SM) with an initial decay time $\tau=2\Delta U_1/J_{U_1}$. Factor 2 appears because the out-flow of energy is only by one wall.}

\label{fig2}

\end{figure}

 In all the simulations we set the flux constant for the system and subsystems i.e.  $J_U=J_{U_1}+J_{U_2}$. We also set the same $J_U$  for all the modes of energy transfer into the system. Thus we make a comparison between different cases using only one parameter, i.e., the energy $\Delta U$ stored in a system. 
  Molecular Dynamics (MD) simulations~\cite{allen1987,verlet1967} (Fig.\ref{fig1}) were performed in the system of fixed number $N = 266240$ of Lennard-Jones (LJ) atoms.  The simulation box was divided into two subsystems (1) and (2) of sizes $L_1$ and $L_2$, respectively. The total size $L_1+L_2$ of the box was constant. 
     Energy was added only to the regions (1),(2)  (Fig.\ref{fig1}) once per 10 time steps in three manners (i),(ii),(iii) as described for the ideal gas model.  When the flux of energy was proportional to the density each LJ atom {received} the same amount of kinetic energy. For flux proportional to temperature  the portion of energy added to each atom was proportional to its kinetic energy.  For energy flux proportional to  volume  the same amount of energy was added to the same volume, i.e.,  all the atoms in a given volume received a given amount of energy equally shared between them. 
All temperature and density profiles in the stationary states are shown in SM for the system with and without internal walls. 
In Fig. \ref{fig2}a we show the dependence of the energy $\Delta U$ stored in the system,  as a function of the size $L_1$ of one subsystem  for fixed fluxes $J_U=J_{U_1}+J_{U_2}$. This figure contains results  for all three scenarios of the energy transfer. All physical quantities are given in LJ units. When $L_1=L_2$, $\Delta U$ reaches a minimal value (equal to the value obtained for the unconstrained system) as expected from Eq(\ref{eq:7}). In Fig(\ref{fig2}b) we show the scaling of the energy per particle stored in the subsystem (1) as a function of the size of the subsystem $L_1$. Finally we observe that after the shutdown of energy flux into the system, the energy decreases as $\exp(-t/\tau)$ (see SM) at short times, with a decay time $\tau=\Delta U/J_U$ (Fig(\ref{fig2}c)).
{In summary,}  the MD simulations of LJ fluid confirm Eq.~(\ref{eq:7}). In most of our simulations {the system was in a single phase state,} but we also confirmed Eq.~(\ref{eq:7})  for the two-phase non-equilibrium system.  We observed a  spontaneous phase separation  and the two phases present in the stationary state with a liquid at the cold boundary and a heated gas inside the simulation box.
In these simulations we could not obtain  stationary states with convection (as in the Rayleigh-Benard cell) irrespective of the value of the energy flux input into the system.
  
\begin{figure}[ht]
\includegraphics[width=0.7\textwidth]{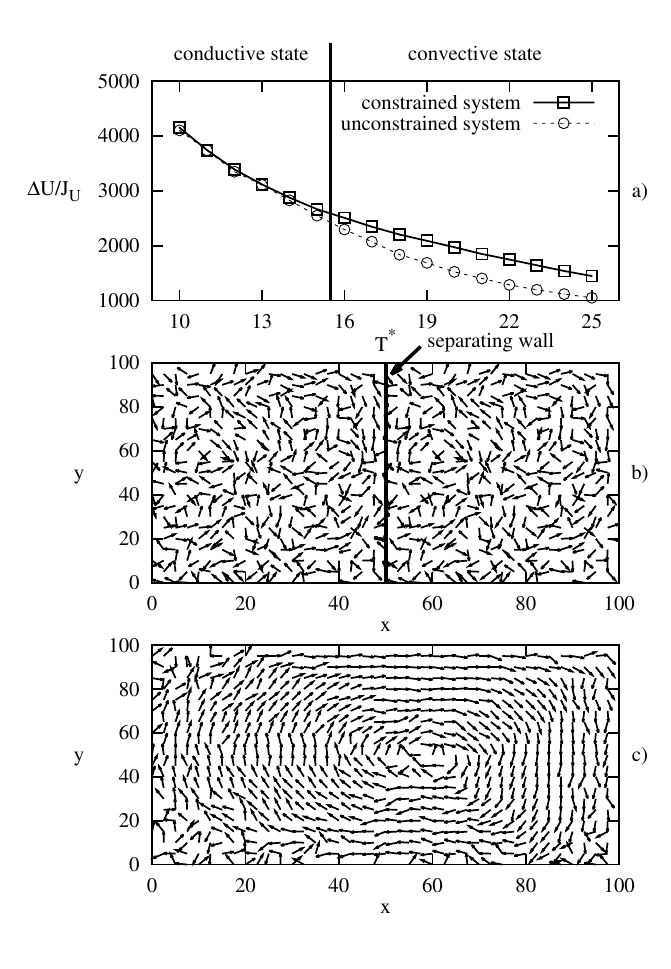} 	
\caption{(a)  $\Delta U/J_U$ as a function of the lower plate dimensionless temperature, $T^*$ in the Rayleigh-Benard 2D system. The vertical solid line marks Rayleigh-Benard instability at about  $T^*=T_{RB}^*=15.5.$ The open squares are the MD data for the constrained system in the conductive state (for all $T^*$) with wall in the middle (see (b) ). The open circles are the MD data for the unconstrained system in the conductive state (see (b)) for $T<T_{RB}$ and convective state (see (c))  for $T>T_{RB}$.  (b) Snapshots of the stationary velocity field (at $T^*=17$) in the conductive state (constrained state) (c) Snapshots of the stationary velocity field in the convective state, stable for $T^*>T_{RB}^*$}.
\label{fig3}
\end{figure}

{\it  2D Rayleigh Benard system of hard discs (HD):} We further tested Eq.~(\ref{eq:7}) in the Rayleigh-Benard (RB) cell (Fig(\ref{fig3}) for  two competing  stationary states. One is called the conductive state Fig(\ref{fig3}b) and second one the convective state as shown in Fig(\ref{fig3}c). For small temperature gradients the conductive state is stable, while for large temperature gradients it is  the convective state that is stable. We use the internal walls to stabilize the conductive state above the transition to the convective state. This methodology allows for comparison of $\Delta U/J_U$ for both states at the same gravitational field and temperatures at the  walls.

Discs of diameter $\sigma$ and mass $m$, were placed in 
a rectangular simulation box of size $L_x d \times L_y d$ (where $d=\sigma/\sqrt{0.4}$). Two sizes were studied $50 d \times 100 d $ (the small system) and $100 d\times 100 d$ (the large system). The dimensionless density, $\rho^*=N\sigma^2/(L_x L_yd^2)=0.4$. The upper plate had a constant temperature $T_0$, the lower plate was heated to temperatures $T=T^* \times T_0$, with $T^* \ge 1$. All the disks were subjected to a gravitational force $ {F} = m (0, -g)$. The HD fluid was simulated with  event-driven dynamics, where discs followed parabolic trajectories between collisions.The discs collided in an elastic manner with the side walls and  between themselves. The collisions with the upper and lower plates allowed for a transfer the thermal energy to the system. Upon collision with the upper or lower plate the disc velocity was drawn from the Maxwell distribution corresponding to the temperature of 
the plate and the direction was chosen randomly from $0$ to $180$ degree angles with respect to the plate\cite{rapaport2004}.
  The total energy of the system is the sum of the kinetic and potential energy of the discs (details in SM).

For $T^*=1$ the system reaches the thermal equilibrium state with an internal energy $U_0$. For $T^*>1$  energy is being pumped into the system at the bottom plate until the system reaches a stationary state, characterized by the energy stored in the system above its equilibrium value, $\Delta U=U - U_0$.
In the stationary state, this system is expected to conduct heat in a conductive manner up to a temperature which corresponds to the RB instability\cite{cross1993}, $T^*_{RB}$. 
This temperature ($T^*_{RB}$) marks the onset of the convective mode of heat transfer,  
characterised by the formation of convective rolls. In our system 
measuring $100d \times 100d$, the transition temperature  $T^*_{RB}$ is about 15.5. For $T^* > 15.5$ we observe a convective state with a single roll filling almost the whole area of the simulation box. However for the system measuring $50d \times 100d$, the system is too small to develop rolls at all the studied temperatures (up to $T^*=25$).  This observation makes it possible to constrain the system by the internal wall and stabilize the conductive state against convective instability. We introduced a constraint into the system by inserting a vertical adiabatic wall in the middle of the $100d \times 100d$ simulation box. The wall divides the system  into two $50d \times 100d$ independent sub-systems (1) and (2). The  $50d \times 100d$ sub-system is too small to develop even a single convective  roll for $ T^*  > 15.5$. Thus a conductive stationary state is observed for each $T^*$ that is considered. Next, by removing the adiabatic wall for $ T^* > 15.5$ we merge two sub-systems in the conductive state into a single 
$100d \times 100d$ system in the convective state. For $T<T_{RB}^*$, the  merging sub-systems  (1) and (2) does not change the conductive stationary state and in particular 
$\Delta U_1+\Delta U_2=\Delta U$ for $J_{U_1}+J_{U_2}=J_U$ at all temperatures $T^*<T_{RB}^*$.

Figure (\ref{fig3}) shows the results of the simulations. The conductive state (shown in Fig(\ref{fig3}b))  is stable for $T^*<T^*_{RB}$. In this state the whole system and the subsystems satisfy the equations $\Delta U=\Delta U_1+\Delta U_2$ and $J_U=J_{U_1}+J_{U_2}$. However for $T^*>T^*_{RB}$ the sub-system (1) and (2) are still in a conductive state, while the whole system without the constraint reaches a convective state (Fig(\ref{fig3}c)). We find  $\Delta U/J_U<(\Delta U_1+\Delta U_2)/(J_{U_1}+J_{U_2})$  for $T^*>T^*_{RB}$ as predicted by Eq.~(\ref{eq:7}).
 Concluding, in the RB cell 
  $\Delta U/J_U$ is minimized in stationary states. 

{\it Conclusions and further discussion}: We have presented a new methodology for the analysis of nonequilibrium states, based on internal constraints known from equilibrium thermodynamics. We have pointed out the importance of the mode of energy transfer into the system and introduced two new quantities characterizing the non-equilibrium stationary state: $\Delta U$ the excess energy stored in the non-equilibrium state over the equilibrium value and $\tau=\Delta U/J_U$, the characteristic time of energy out-flow from the system after shutdown of energy flux into the system. $\Delta U>0$  in all examples discussed in this paper, because we used a reference equilibrium state of lower energy, than the energy of the stationary state. 

We observed that in all cases studied the quantity $\Delta U/J_U$ is minimized  in stationary states. However, { the following counterexample suggests that this may not be a general principle
\cite{footnote1}.
Consider}  a huge box with adiabatic walls attached to our system by a thin { heat-conducting wire ending  at a point inside our system. The point is chosen in such a way, that the temperature at this point is lower in the constrained system than in the unconstrained one. Such a point always exists in the systems studied.} This box stores an extra energy during the energy flow from our system to the box. The flow stops  when the temperature of the box reaches the { stationary state} temperature in the point of choice in our system. This sort of  box does not influence the stationary state of the system but, nonetheless, changes the total amount of energy stored in the total system. Since in this way {one} an arbitrary large amount of energy can be stored, we cannot claim that for a fixed flux the energy stored in the system is minimized. We can make many different variants of this counterexample, { in particular,} allowing a small flux through this box and in this way 
affecting the final non-equilibrium state of the system. { Since such  situations are not eliminated by the current formulation 
of  prerequisites, their  further analysis is needed. Those prerequisites which are related to non-equilibrium states require special attention. Indeed, the} division of a system into subsystems, so obvious for the equilibrium state is not at all obvious  for non-equilibrium states. This is {due to} the crucial role of surfaces bounding the system { establishing} the final non-equilibrium, stationary state.

Our methodology and observations require further tests. Such tests can be performed for chemical systems with many competing stationary  states or in hydrodynamic systems. One example, which we are going to study is the reaction between nitrogen dioxide and nitrogen tetroxide~\cite{creel1976}. A system consisting of these two chemical compounds is illuminated by light. { Light  is absorbed} by nitogen dioxide but not by nitrogen tetroxide. The absorption of light results in heating { of the sample and in an} increase in the backward reaction from tetroxide to dioxide. In this system many stationary states appear. We hope that such test will further support our current observations. 
 \newpage 

\newpage
\clearpage
\begin{acknowledgments}
*rholyst@ichf.edu.pl. This work was supported by the Maestro grant UMO-2016/22/A/ST4/00017 from the National Science Centre, Poland. The authors gratefully acknowledge the computational grant from the Supercomputing and Networking Centre (PCSS) in Poznan, Poland. 
\end{acknowledgments}


\begin{thebibliography}{100}
\bibitem{footnote} Note that $\Delta U$ in Eq. (4) is positive because the energy is supplied to the system  and the flux $J_U > 0$.
\bibitem{footnote1} This counterexample has been presented to us by the referee.
\end{thebibliography}
\end{document}